\documentclass{svmult}
\usepackage{epsfig,graphicx} 
\usepackage[bottom]{footmisc}
\usepackage{natbib,url}      
\bibpunct{(}{)}{;}{a}{}{,}   
\def\cite#1{\citealp{#1}}    
\def\authorindex#1{}  
\def\figspath{.}      

\begin{document}\newcount\preprintheader\preprintheader=1 


\def\thisvolume{these proceedings}

\def\aj{{AJ}}			
\def\araa{{ARA\&A}}		
\def\apj{{ApJ}}			
\def\apjl{{ApJ}}		
\def\apjs{{ApJS}}		
\def\ao{{Appl.\ Optics}} 
\def\apss{{Ap\&SS}}		
\def\aap{{A\&A}}		
\def\aapr{{A\&A~Rev.}}		
\def\aaps{{A\&AS}}		
\def\an{{Astron.\ Nachrichten}}
\def\aspcs{{ASP Conf.\ Ser.}}
\def\assp{{Astrophys.\ \& Space Sci.\ Procs., Springer, Heidelberg}}
\def\azh{{AZh}}			
\def\baas{{BAAS}}		
\def\jrasc{{JRASC}}	
\def\memras{{MmRAS}}		
\def\mnras{{MNRAS}}
\def\nat{{Nat}}		
\def\pra{{Phys.\ Rev.\ A}} 
\def\prb{{Phys.\ Rev.\ B}}		
\def\prc{{Phys.\ Rev.\ C}}		
\def\prd{{Phys.\ Rev.\ D}}		
\def\prl{{Phys.\ Rev.\ Lett.}} 
\def\pasp{{PASP}}
\def\pasj{{PASJ}}		
\def\qjras{{QJRAS}}
\def\science{{Sci}}		
\def\skytel{{S\&T}}		
\def\solphys{{Solar\ Phys.}} 
\def\sovast{{Soviet\ Ast.}}  
\def\ssr{{Space\ Sci.\ Rev.}}
\def\svassp{{Astrophys.\ Space Sci.\ Procs., Springer, Heidelberg}}
\def\zap{{ZAp}}			
\let\astap=\aap
\let\apjlett=\apjl
\let\apjsupp=\apjs
\def\grl{{Geophys.\ Res.\ Lett.}}  
\def\jgr{{J. Geophys.\ Res.}} 

\def\ion#1#2{{\rm #1}\,{\uppercase{#2}}}  
\def\deg{\hbox{$^\circ$}}
\def\sun{\hbox{$\odot$}}
\def\earth{\hbox{$\oplus$}}
\def\la{\mathrel{\hbox{\rlap{\hbox{\lower4pt\hbox{$\sim$}}}\hbox{$<$}}}}
\def\ga{\mathrel{\hbox{\rlap{\hbox{\lower4pt\hbox{$\sim$}}}\hbox{$>$}}}}
\def\sq{\hbox{\rlap{$\sqcap$}$\sqcup$}}
\def\arcmin{\hbox{$^\prime$}}
\def\arcsec{\hbox{$^{\prime\prime}$}}
\def\fd{\hbox{$.\!\!^{\rm d}$}}
\def\fh{\hbox{$.\!\!^{\rm h}$}}
\def\fm{\hbox{$.\!\!^{\rm m}$}}
\def\fs{\hbox{$.\!\!^{\rm s}$}}
\def\fdg{\hbox{$.\!\!^\circ$}}
\def\farcm{\hbox{$.\mkern-4mu^\prime$}}
\def\farcs{\hbox{$.\!\!^{\prime\prime}$}}
\def\fp{\hbox{$.\!\!^{\scriptscriptstyle\rm p}$}}
\def\micron{\hbox{$\mu$m}}
\def\onehalf{\hbox{$\,^1\!/_2$}}	
\def\onethird{\hbox{$\,^1\!/_3$}}
\def\twothirds{\hbox{$\,^2\!/_3$}}
\def\onequarter{\hbox{$\,^1\!/_4$}}
\def\threequarters{\hbox{$\,^3\!/_4$}}
\def\ubv{\hbox{$U\!BV$}}		
\def\ubvr{\hbox{$U\!BV\!R$}}		
\def\ubvri{\hbox{$U\!BV\!RI$}}		
\def\ubvrij{\hbox{$U\!BV\!RI\!J$}}		
\def\ubvrijh{\hbox{$U\!BV\!RI\!J\!H$}}		
\def\ubvrijhk{\hbox{$U\!BV\!RI\!J\!H\!K$}}		
\def\ub{\hbox{$U\!-\!B$}}		
\def\bv{\hbox{$B\!-\!V$}}		
\def\vr{\hbox{$V\!-\!R$}}		
\def\ur{\hbox{$U\!-\!R$}}


\def\labelitemi{{\bf --}}  

\def\rmit#1{{\it #1}}              
\def\rmit#1{{\rm #1}}              
\def\etal{\rmit{et al.}}           
\def\etc{\rmit{etc.}}           
\def\ie{\rmit{i.e.,}}              
\def\eg{\rmit{e.g.,}}              
\def\cf{cf.}                       
\def\viz{\rmit{viz.}}
\def\vs{\rmit{vs.}}

\def\rot{\hbox{\rm rot}}
\def\div{\hbox{\rm div}}
\def\lesssim{\mathrel{\hbox{\rlap{\hbox{\lower4pt\hbox{$\sim$}}}\hbox{$<$}}}}
\def\gtrsim{\mathrel{\hbox{\rlap{\hbox{\lower4pt\hbox{$\sim$}}}\hbox{$>$}}}}
\def\mathstacksym#1#2#3#4#5{\def#1{\mathrel{\hbox to 0pt{\lower 
    #5\hbox{#3}\hss} \raise #4\hbox{#2}}}}
\mathstacksym\lesssim{$<$}{$\sim$}{1.5pt}{3.5pt} 
\mathstacksym\gtrsim{$>$}{$\sim$}{1.5pt}{3.5pt} 
\mathstacksym\lrarrow{$\leftarrow$}{$\rightarrow$}{2pt}{1pt} 
\mathstacksym\lessgreat{$>$}{$<$}{3pt}{3pt} 

\def\dif{\: {\rm d}}                       
\def\ep{\:{\rm e}^}                        
\def\dash{\hbox{$\,-\,$}}                  
\def\is{\!=\!}                             

\def\starname#1#2{${#1}$\,{\rm {#2}}}  
\def\Teff{\hbox{$T_{\rm eff}$}}   

\def\kms{\hbox{km$\;$s$^{-1}$}}
\def\ms{\hbox{m$\;$s$^{-1}$}}
\def\Mxcm{\hbox{Mx\,cm$^{-2}$}}    

\def\Bapp{\hbox{$B_{\rm app}$}}    

\def\komega{($k, \omega$)}                 
\def\kf{($k_h,f$)}                         
\def\VminI{\hbox{$V\!\!-\!\!I$}}           
\def\IminI{\hbox{$I\!\!-\!\!I$}}           
\def\VminV{\hbox{$V\!\!-\!\!V$}}           
\def\Xt{\hbox{$X\!\!-\!t$}}                

\def\level #1 #2#3#4{$#1 \: ^{#2} \mbox{#3} ^{#4}$}   

\def\specchar#1{\uppercase{#1}}    
\def\AlI{\mbox{Al\,\specchar{i}}}  
\def\BI{\mbox{B\,\specchar{i}}} 
\def\BII{\mbox{B\,\specchar{ii}}}  
\def\BaI{\mbox{Ba\,\specchar{i}}}  
\def\BaII{\mbox{Ba\,\specchar{ii}}} 
\def\CI{\mbox{C\,\specchar{i}}} 
\def\CII{\mbox{C\,\specchar{ii}}} 
\def\CIII{\mbox{C\,\specchar{iii}}} 
\def\CIV{\mbox{C\,\specchar{iv}}} 
\def\CaI{\mbox{Ca\,\specchar{i}}} 
\def\CaII{\mbox{Ca\,\specchar{ii}}} 
\def\CaIII{\mbox{Ca\,\specchar{iii}}} 
\def\CoI{\mbox{Co\,\specchar{i}}} 
\def\CrI{\mbox{Cr\,\specchar{i}}} 
\def\CriI{\mbox{Cr\,\specchar{ii}}} 
\def\CsI{\mbox{Cs\,\specchar{i}}} 
\def\CsII{\mbox{Cs\,\specchar{ii}}} 
\def\CuI{\mbox{Cu\,\specchar{i}}} 
\def\FeI{\mbox{Fe\,\specchar{i}}} 
\def\FeII{\mbox{Fe\,\specchar{ii}}} 
\def\FeIX{\mbox{Fe\,\specchar{ix}}}
\def\FeX{\mbox{Fe\,\specchar{x}}}
\def\FeXVI{\mbox{Fe\,\specchar{xvi}}}
\def\FrI{\mbox{Fr\,\specchar{i}}}
\def\HI{\mbox{H\,\specchar{i}}} 
\def\HII{\mbox{H\,\specchar{ii}}} 
\def\Hmin{\hbox{\rmH$^{^{_{\scriptstyle -}}}$}}      
\def\Hemin{\hbox{{\rm He}$^{^{_{\scriptstyle -}}}$}} 
\def\HeI{\mbox{He\,\specchar{i}}} 
\def\HeII{\mbox{He\,\specchar{ii}}} 
\def\HeIII{\mbox{He\,\specchar{iii}}} 
\def\KI{\mbox{K\,\specchar{i}}} 
\def\KII{\mbox{K\,\specchar{ii}}} 
\def\KIII{\mbox{K\,\specchar{iii}}} 
\def\LiI{\mbox{Li\,\specchar{i}}} 
\def\LiII{\mbox{Li\,\specchar{ii}}} 
\def\LiIII{\mbox{Li\,\specchar{iii}}} 
\def\MgI{\mbox{Mg\,\specchar{i}}} 
\def\MgII{\mbox{Mg\,\specchar{ii}}} 
\def\MgIII{\mbox{Mg\,\specchar{iii}}} 
\def\MnI{\mbox{Mn\,\specchar{i}}} 
\def\NI{\mbox{N\,\specchar{i}}}
\def\NIV{\mbox{N\,\specchar{iv}}}
\def\NaI{\mbox{Na\,\specchar{i}}}
\def\NaII{\mbox{Na\,\specchar{ii}}}
\def\NaIII{\mbox{Na\,\specchar{iii}}}
\def\NeVIII{\mbox{Ne\,\specchar{viii}}} 
\def\NiI{\mbox{Ni\,\specchar{i}}} 
\def\NiII{\mbox{Ni\,\specchar{ii}}}
\def\NiIII{\mbox{Ni\,\specchar{iii}}} 
\def\OI{\mbox{O\,\specchar{i}}} 
\def\OVI{\mbox{O\,\specchar{vi}}}
\def\RbI{\mbox{Rb\,\specchar{i}}} 
\def\SII{\mbox{S\,\specchar{ii}}} 
\def\SiI{\mbox{Si\,\specchar{i}}} 
\def\SiII{\mbox{Si\,\specchar{ii}}} 
\def\SrI{\mbox{Sr\,\specchar{i}}}
\def\SrII{\mbox{Sr\,\specchar{ii}}}
\def\TiI{\mbox{Ti\,\specchar{i}}} 
\def\TiII{\mbox{Ti\,\specchar{ii}}} 
\def\TiIII{\mbox{Ti\,\specchar{iii}}} 
\def\TiIV{\mbox{Ti\,\specchar{iv}}} 
\def\VI{\mbox{V\,\specchar{i}}} 
\def\HtwoO{\mbox{H$_2$O}}        
\def\Otwo{\mbox{O$_2$}}          

\def\Halpha{\mbox{H\hspace{0.1ex}$\alpha$}} 
\def\Ha{\mbox{H\hspace{0.2ex}$\alpha$}}
\def\Hbeta{\mbox{H\hspace{0.2ex}$\beta$}}
\def\Hgamma{\mbox{H\hspace{0.2ex}$\gamma$}}
\def\Hdelta{\mbox{H\hspace{0.2ex}$\delta$}}
\def\Hepsilon{\mbox{H\hspace{0.2ex}$\epsilon$}}
\def\Hzeta{\mbox{H\hspace{0.2ex}$\zeta$}}
\def\Lyalpha{\mbox{Ly$\hspace{0.2ex}\alpha$}}
\def\Lybeta{\mbox{Ly$\hspace{0.2ex}\beta$}}
\def\Lygamma{\mbox{Ly$\hspace{0.2ex}\gamma$}}
\def\Lycont{\mbox{Ly\hspace{0.2ex}{\small cont}}}
\def\Baalpha{\mbox{Ba$\hspace{0.2ex}\alpha$}}
\def\Babeta{\mbox{Ba$\hspace{0.2ex}\beta$}}
\def\Bacont{\mbox{Ba\hspace{0.2ex}{\small cont}}}
\def\Paalpha{\mbox{Pa$\hspace{0.2ex}\alpha$}}
\def\Bralpha{\mbox{Br$\hspace{0.2ex}\alpha$}}

\def\NaD{\mbox{Na\,\specchar{i}\,D}}    
\def\NaDone{\mbox{Na\,\specchar{i}\,\,D$_1$}}
\def\NaDtwo{\mbox{Na\,\specchar{i}\,\,D$_2$}}
\def\NaID{\mbox{Na\,\specchar{i}\,\,D}}
\def\NaIDone{\mbox{Na\,\specchar{i}\,\,D$_1$}}
\def\NaIDtwo{\mbox{Na\,\specchar{i}\,\,D$_2$}}
\def\Done{\mbox{D$_1$}}
\def\Dtwo{\mbox{D$_2$}}

\def\Mgbone{\mbox{Mg\,\specchar{i}\,b$_1$}}
\def\Mgbtwo{\mbox{Mg\,\specchar{i}\,b$_2$}}
\def\Mgbthree{\mbox{Mg\,\specchar{i}\,b$_3$}}
\def\MgIb{\mbox{Mg\,\specchar{i}\,b}}
\def\MgIbone{\mbox{Mg\,\specchar{i}\,b$_1$}}
\def\MgIbtwo{\mbox{Mg\,\specchar{i}\,b$_2$}}
\def\MgIbthree{\mbox{Mg\,\specchar{i}\,b$_3$}}

\def\CaIIK{\mbox{Ca\,\specchar{ii}\,K}}       
\def\CaIIH{\mbox{Ca\,\specchar{ii}\,H}}
\def\CaIIHK{\mbox{Ca\,\specchar{ii}\,H\,\&\,K}}
\def\HK{\mbox{H\,\&\,K}}
\def\Kthree{\mbox{K$_3$}}      
\def\Hthree{\mbox{H$_3$}}
\def\Ktwo{\mbox{K$_2$}}
\def\Htwo{\mbox{H$_2$}}
\def\Kone{\mbox{K$_1$}}     
\def\Hone{\mbox{H$_1$}}     
\def\KtwoV{\mbox{K$_{2V}$}}
\def\KtwoR{\mbox{K$_{2R}$}}
\def\KoneV{\mbox{K$_{1V}$}}
\def\KoneR{\mbox{K$_{1R}$}}
\def\HtwoV{\mbox{H$_{2V}$}}
\def\HtwoR{\mbox{H$_{2R}$}}
\def\HoneV{\mbox{H$_{1V}$}}
\def\HoneR{\mbox{H$_{1R}$}}

\def\hk{\mbox{h\,\&\,k}}
\def\kthree{\mbox{k$_3$}}    
\def\hthree{\mbox{h$_3$}}
\def\ktwo{\mbox{k$_2$}}
\def\htwo{\mbox{h$_2$}}
\def\kone{\mbox{k$_1$}}     
\def\hone{\mbox{h$_1$}}     
\def\ktwoV{\mbox{k$_{2V}$}}
\def\ktwoR{\mbox{k$_{2R}$}}
\def\koneV{\mbox{k$_{1V}$}}
\def\koneR{\mbox{k$_{1R}$}}
\def\htwoV{\mbox{h$_{2V}$}}
\def\htwoR{\mbox{h$_{2R}$}}
\def\honeV{\mbox{h$_{1V}$}}
\def\honeR{\mbox{h$_{1R}$}}

\ifnum\preprintheader=1     
\makeatletter  
\def\@maketitle{\newpage
\markboth{}{}%
  {\mbox{} \vspace*{-8ex} \par 
   \em \footnotesize To appear in ``Magnetic Coupling between the Interior 
       and the Atmosphere of the Sun'', eds. S.~S.~Hasan and R.~J.~Rutten, 
       Astrophysics and Space Science Proceedings, Springer-Verlag, 
       Heidelberg, Berlin, 2009.} \vspace*{-5ex} \par
 \def\lastand{\ifnum\value{@inst}=2\relax
                 \unskip{} \andname\
              \else
                 \unskip \lastandname\
              \fi}%
 \def\and{\stepcounter{@auth}\relax
          \ifnum\value{@auth}=\value{@inst}%
             \lastand
          \else
             \unskip,
          \fi}%
  \raggedright
 {\Large \bfseries\boldmath
  \pretolerance=10000
  \let\\=\newline
  \raggedright
  \hyphenpenalty \@M
  \interlinepenalty \@M
  \if@numart
     \chap@hangfrom{}
  \else
     \chap@hangfrom{\thechapter\thechapterend\hskip\betweenumberspace}
  \fi
  \ignorespaces
  \@title \par}\vskip .8cm
\if!\@subtitle!\else {\large \bfseries\boldmath
  \vskip -.65cm
  \pretolerance=10000
  \@subtitle \par}\vskip .8cm\fi
 \setbox0=\vbox{\setcounter{@auth}{1}\def\and{\stepcounter{@auth}}%
 \def\thanks##1{}\@author}%
 \global\value{@inst}=\value{@auth}%
 \global\value{auco}=\value{@auth}%
 \setcounter{@auth}{1}%
{\lineskip .5em
\noindent\ignorespaces
\@author\vskip.35cm}
 {\small\institutename\par}
 \ifdim\pagetotal>157\p@
     \vskip 11\p@
 \else
     \@tempdima=168\p@\advance\@tempdima by-\pagetotal
     \vskip\@tempdima
 \fi
}
\makeatother     
\fi

\title*{CME Observations from STEREO}

\author{N. Srivastava}

\authorindex{Srivastava, N.}

\institute{Udaipur Solar Observatory, Physical Research Laboratory, Udaipur}

\maketitle

\setcounter{footnote}{0}  

\begin{abstract} 
  Coronal mass ejections (CMEs) are spectacular ejections of material
  from the Sun as seen in the coronal field of view.  Regular observations are possible with
  both ground-based and space-based coronagraphs. I present our
  current understanding of CMEs based on multi-wavelength observations
  from groundbased instruments as well as from space missions such as
  SoHO.  Based on the continuous and multi-wavelength observations of
  CMEs from SoHO over a period of more than a solar cycle, the
  physical properties of CMEs are described. Recent observations of
  CMEs with the SECCHI coronagraphs, namely COR1 and COR2 aboard the
  twin STEREO spacecrafts A and B are also presented.  STEREO
  surpasses previous missions by providing a 3-D view of CME structure
  from two vantage points.  Applications of STEREO observations to 3-D
  reconstructions of the leading edge of CMEs are described.

\end{abstract}
\section{Introduction}      \label{rutten-sec:introduction}

Coronal mass ejections (CMEs) were first observed by OSO-7 in 1971
(\cite{Tousey73}). Since then, a number of space-based and
ground-based coronagraphs have been regularly recording images of the
corona and CMEs.  These observations have led to a fairly good general
understanding of CMEs.  They were traditionally observed in white
light or continuum images by space-based coronagraphs such as Solwind,
the Solar Maximum Mission (SMM), and LASCO/C2\&C3, and with
ground-based instruments such as the MK\,III and MK\,IV coronagraphs
at Mauna Loa. However, CMEs have also been recorded in coronal
emission lines  for e.g., Fe\,XIV or Fe\,X, such as with the Norikura
observatory and with LASCO-C1 onboard SoHO.
Table~\ref{n-srivastava-tbl:comparison} gives the field of view of
various coronagraphs. SoHO was the first successful space mission with
multiple instruments onboard to record various aspects of transient
activity at multiple wavelengths. The Extreme Ultraviolet telescope
(EIT) onboard SoHO not only can track down the source region of a CME,
but SoHO also can record CME propagation through a field of view of
1--32~R$_{\odot}$ with the Large Angle Spectrometric COronagraphs
(LASCO) at an improved time cadence (\cite{Delaboudiniere95};
\cite{Brueckner95}). These continuous, multi-wavelength, and high
time-cadence images have led to a better understanding of CME initiation
and propagation.  The details are presented in various reviews
(\cite{2007ISSI}; \cite{Schwenn07}).

An important research area in which SoHO observations proved to be
extremely useful is space weather prediction.  A large number of
studies have been undertaken in this direction, in particular, for
predicting the arrival time of CMEs at the Earth through continuous
monitoring of the Sun and CME tracking. This helps in predicting the
time of commencement of geomagnetic storms, which are normally
expected to occur when a high-speed CME is directed towards the Earth
and its magnetic field is oriented southward so as to ensure an
effective coupling between the Earth's magnetic field and the
propagating magnetic cloud. However, most prediction schemes used
plane-of-sky speeds as proxies of radial speeds, which has led to
large errors in the estimated arrival times (\cite{Schwenn05}). This
error can be attributed to the lack of capability of SoHO to measure
the radial speeds of CMEs. The problem is expected to be overcome with
new observations from STEREO, which comprises twin spacecrafts with
identical sets of instruments (\cite{Howard08}). The main goal of
STEREO is to improve our understanding of the 3-D structure of CMEs
with accurate and direct estimations of their true speeds and
propagation directions.

\begin{table}
\begin{center}
\renewcommand{\arraystretch}{1.1}
\caption[]{\label{n-srivastava-tbl:comparison}{Comparison of different space-based coronagraphs used to record CMEs}}
\begin{tabular*}{0.9\textwidth}{@{\extracolsep{\fill}}cccc}
\hline
Coronagraph	&Year	& Field of view	&Resolution\\
\hline
OSO-7	&1971--1973	&3.0--10.0 R$_{\odot}$&	3\arcmin\\[0.7ex]
Skylab&	1973--1974&	2.0--6.0 R$_{\odot}$&	5\arcsec\\[0.7ex]
Solwind/P78-I &	1979--1985 &	3.0--10.0 R$_{\odot}$ &	Same as OSO\\[0.7ex]
SMM &	1980	&1.6--6.0 R$_{\odot}$ &	30\arcsec\\
& 1984--1989 & &\\[0.7ex]
LASCO &	& &\\
C1& 1995--1998&1.1--3.0 R$_{\odot}$ & 11.2\arcsec\\
C2& 1995--&2.0--6.0 R$_{\odot}$&23.2\arcsec\\ 
C3&& 3.7--32.0 R$_{\odot}$ &	112\arcsec\\[0.7ex]
SECCHI	& 2006--& &\\
COR1 & &1.4--4.0 R$_{\odot}$ & 7.5\arcsec\\
COR2& &2.0--15.0 R$_{\odot}$ & 14\arcsec\\
\hline
\end{tabular*}
\end{center}
\end{table}

In what follows, statistical properties of CMEs are highlighted on the
basis of SoHO observations and new observations from the SECCHI
coronagraphs onboard the STEREO mission are discussed. In particular,
I showcase the results of 3-D reconstruction of a CME leading edge,
using SECCHI-COR1 and COR2 observations, and discuss implications for
space-weather prediction.

\begin{figure}
  \sidecaption
  \includegraphics[width=5cm]{\figspath/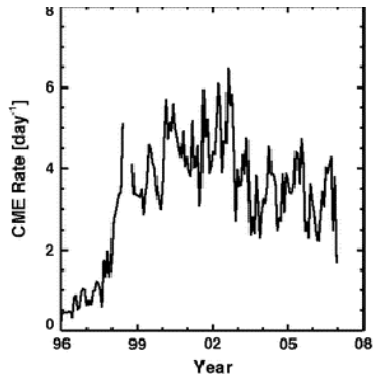}
\caption[]{\label{n-srivastava-fig:cme-rate}  
  The rate of occurrence of LASCO CMEs averaged over a full solar
  cycle, \ie\ 1996--2007 (adapted from \cite{Gopalswamy09}).}
\end{figure}

\section{What have we learnt from LASCO observations?}

The three LASCO coronagraphs (\cite{Brueckner95}) onboard SoHO provide
a nested field of view and have tracked a large number of CMEs from
their launch at the solar surface. These observations confirm that
CMEs generally have a three-part structure. They have a bright leading
edge, a dark cavity believed to have a high magnetic field, followed
by a bright and intense knot which mainly comprises prominence
material. A large number of earthward-directed CMEs were recorded by
LASCO. These are observed as full halos in which brightness
enhancement is seen in an angular span of 360$^{\circ}$ around the
occulter, while in the case of a partial halo, the brightness is seen
over an angular span of more than 120$^{\circ}$. The kinematics of
full and partial halos have also been studied (\cite{Wang02};
\cite{Zhang03}; \cite{Zhao03}; \cite{dal Lago04}). These studies were
aimed at inferring the travel time of the CMEs to the Earth.  The
estimated arrival time of a CME at the Earth is an important input for
forecasting the time of occurrence of the resulting geomagnetic storms,
provided there is a strong southward component of the propagating
magnetic cloud.

\begin{figure}  
  \centering
  \includegraphics[width=2.35in]{\figspath/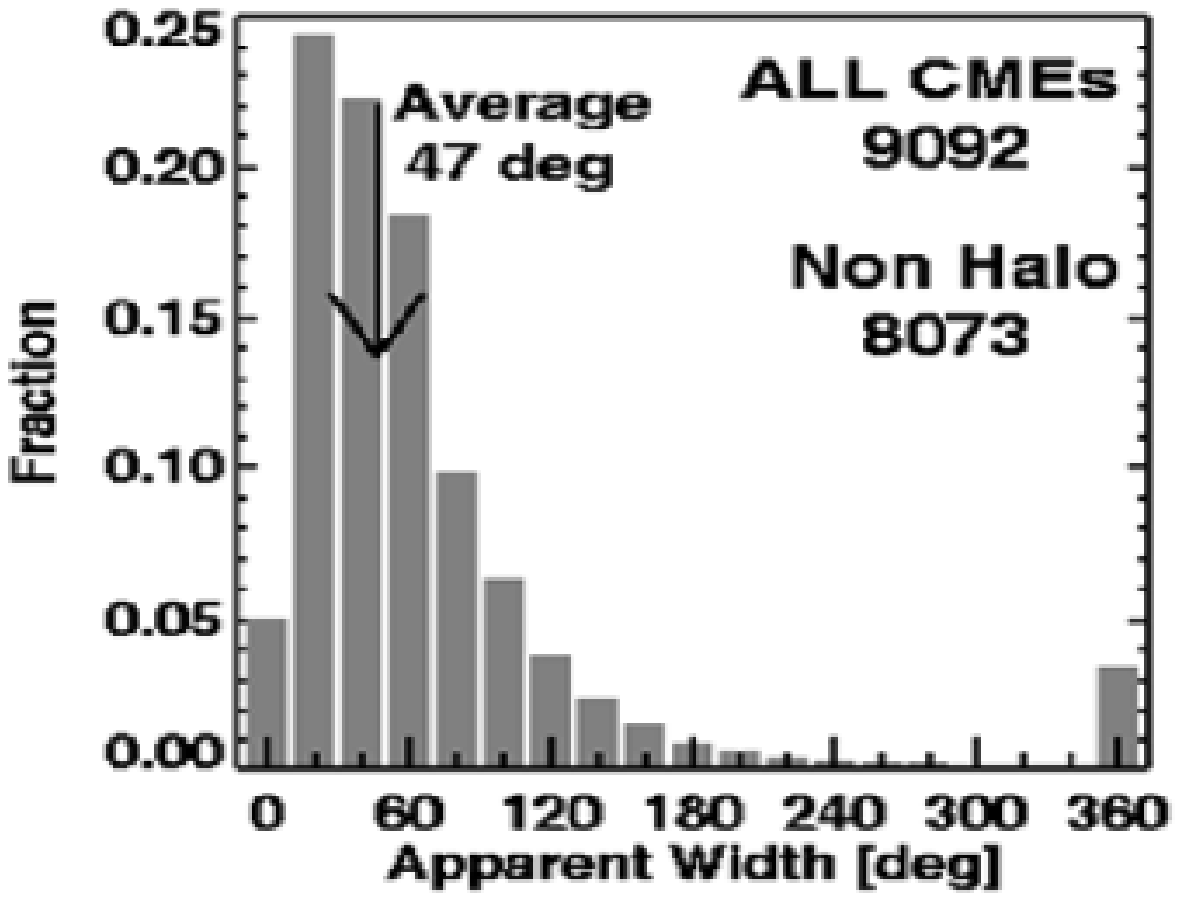}
  \includegraphics[width=2.20in]{\figspath/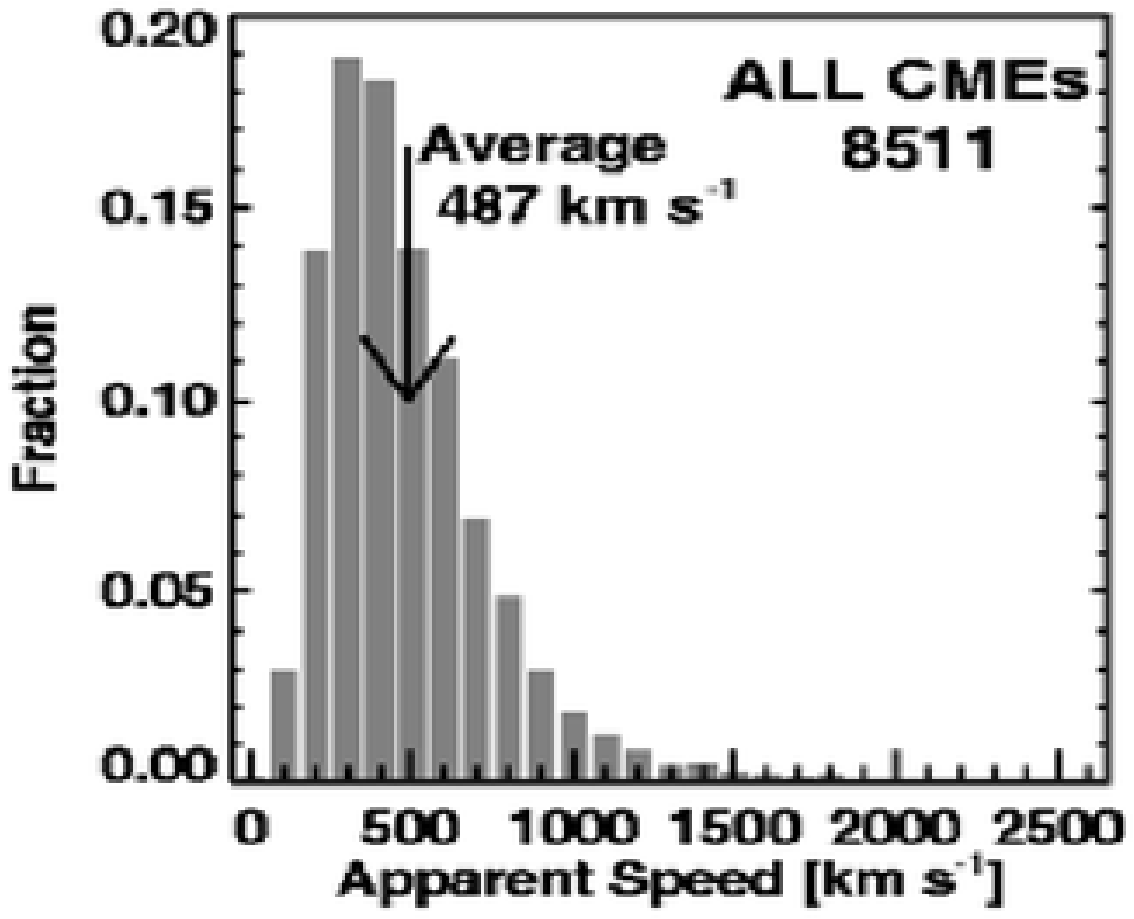}
\caption[]{\label{n-srivastava-fig:width-speed} 
  {\em Left\/}: average width of all non-halo (angular width
  0--120$^{\circ}$) LASCO CMEs. {\em Right\/}: the projected speed
  of both halo and limb CMEs as recorded by the LASCO coronagraphs.}
\end{figure}

A huge dataset has been collected with the LASCO coronagraphs over
more than a solar cycle.  It allows us to study CME properties over a
large time period.  The rate of occurrence of CMEs has been found to
be 0.3/day during the solar minimum around 1996, and it slowly
increased to 5--6~CMEs/day during the solar maxima
(Fig.~\ref{n-srivastava-fig:cme-rate}). \citet{Yashiro04} studied
the variation of angular or apparent width of all CMEs that occurred
during 1996--2003 and found that all non-halo CMEs have an average
angular width of 47$^{\circ}$ (left-hand panel of
Fig.~\ref{n-srivastava-fig:width-speed}). The kinematics of the CMEs
recorded by LASCO have also been studied by
\citet{Yashiro04}. They measured the projected plane-of-sky speeds of
CMEs and found that these lie in the range 10--3000~\kms, the average
being 487~\kms (right-hand panel in Fig.~\ref{n-srivastava-fig:width-speed}). They also found that the average speed varies with the
solar cycle, the average plane-of-sky speeds in the descending phase
being lower than in the minimum.

\begin{figure}  
  \sidecaption
  \includegraphics[width=0.6\textwidth]{\figspath/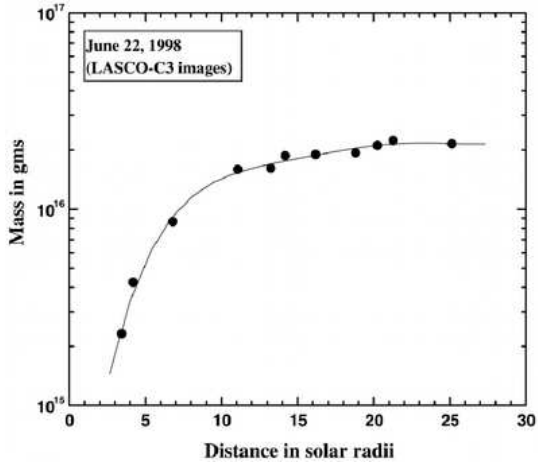}
\caption[]{\label{n-srivastava-fig:mass}  
  The variation of total mass of the CME observed on June 21--22, 1998,
  with distance to the center of the Sun (adapted from
  \cite{Srivastava00}).}
\end{figure}

The white-light coronal images have also been used to measure the
density of CMEs by estimating their excess brightness. The density
values thus obtained were used to derive the mass of CMEs, which
ranges between $10^{15}$--$10^{16}$~g.  The average CME mass is
smaller than the pre-SoHO values because LASCO could record low-mass
CMEs owing to its high sensitivity. Usually, most CMEs show an initial
increase in mass before reaching a constant value, as shown in
Fig.~\ref{n-srivastava-fig:mass}.  \citet{Yashiro04} found that the
mass estimates are uncertain by a factor of two.  Using the estimated
values of density and the measured values of the plane-of-sky speeds
of the CMEs, the kinetic energies of CMEs were found to lie in the
range 10$^{31}$--10$^{32}$~erg. A comparison of statistical properties
of white-light CMEs recorded by LASCO and recorded by earlier
space-borne coronagraphs, namely, Solwind and SMM, is given in
Table~\ref{n-srivastava-tbl:properties}.

\begin{table}
\begin{center}
\renewcommand{\arraystretch}{1.1}
\caption[]{\label{n-srivastava-tbl:properties}{Average CME properties}}
\begin{tabular*}{0.8\textwidth}{@{\extracolsep{\fill}}lrr}
\hline
Parameter	& LASCO	& Solwind/SMM\\
\hline
observing duty cycle	&81.7\%	&66.5\%\\
kinetic energy [erg]&	$2.6 \times 10^{30}$ &  $3.5\times10^{30}$\\
average mass [gm]&	$1.4 \times 10^{15}$&	$4.1\times 10^{15}$\\
mass flux [gm/day]& $2.7 \times10^{15}$	&$7.5 \times10^{15}$ \\
average speed [km s$^{-1}$]& 487 & 349 \\
speed range [km s$^{-1}$]& $10-3000$   & $80-1042$\\
rate of occurrence [CME/day] &  &\\
~~cycle minimum & $0.31-0.77$ & 0.5 \\
~~cycle maximum & $1.75-3.11$ & $5-6$\\
angular width [$^\circ$] &47 & 40\\
\hline
\end{tabular*}
\end{center}
\end{table}

A number of studies have been made on height-time profiles of various
types of CMEs, indicating that there is broad spectrum of
profiles. At one end of the speed spectrum, there are CMEs which rise
gradually with slow speed (50--100~\kms) over a period of
several hours. They reach a terminal speed of 300--400~\kms,
equivalent to that of the slow solar wind speed at 20~R$_{\odot}$
(\cite{Srivastava99}; \cite{Srivastava00}). These are generally
associated with eruptive filaments or prominences. At the other end of
the speed spectrum are fast events, with speeds greater than 600~\kms\
which undergo maximum acceleration in the lower corona. Such
CMEs are generally associated with flares (\cite{Zhang04}).

\begin{figure}  
  \centering
  \includegraphics[width=0.48\textwidth]{\figspath/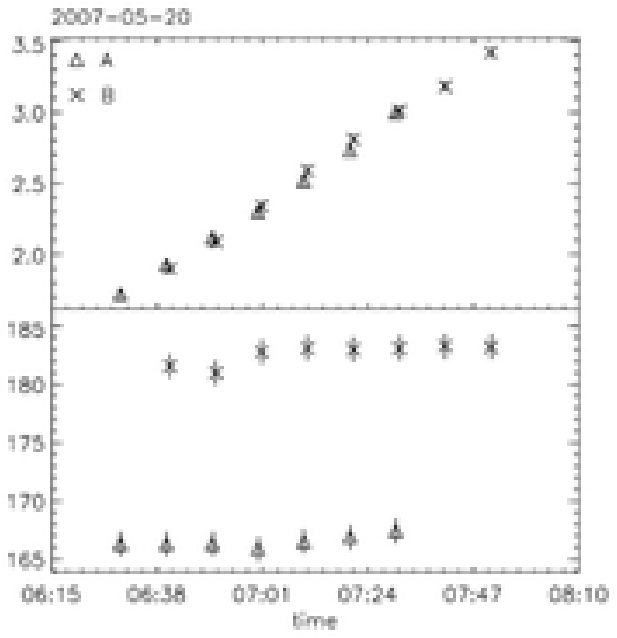}
  \includegraphics[width=0.48\textwidth]{\figspath/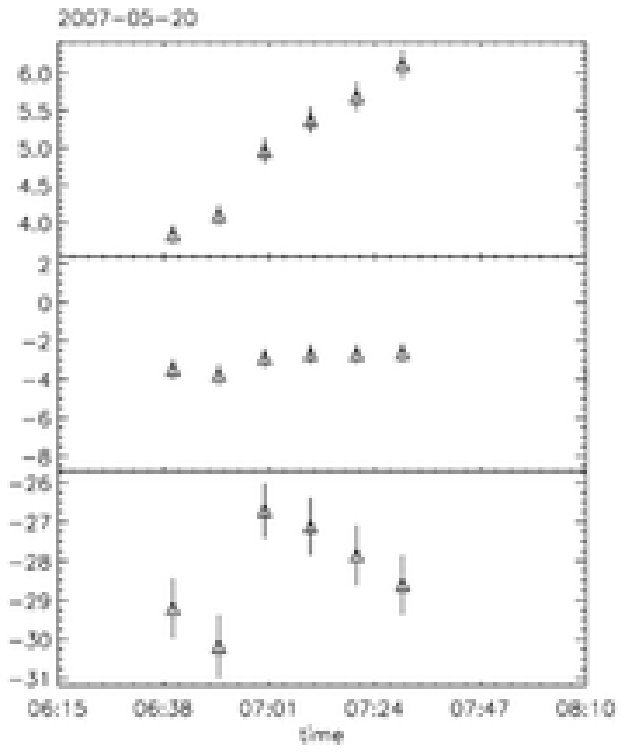}
\caption[]{\label{n-srivastava-fig:mierla-httime}  
  {\em Left\/}: projected height (top) and position
  angle (bottom) of an identifiable feature along the leading edge of
  the 20 May 2007 CME, as observed in the COR1A and B
  coronagraphs. The position angle of the selected feature is measured
  counterclockwise from the North ecliptic pole.  {\em Right\/}:
  true height (top), ecliptic longitude (middle), and ecliptic
  latitude (bottom) for the same feature, obtained using the
  height-time technique (adapted from \cite{Mierla08}).}
\end{figure}

\begin{figure}
\begin{tabular}{cc}
\includegraphics[width=5.8cm]{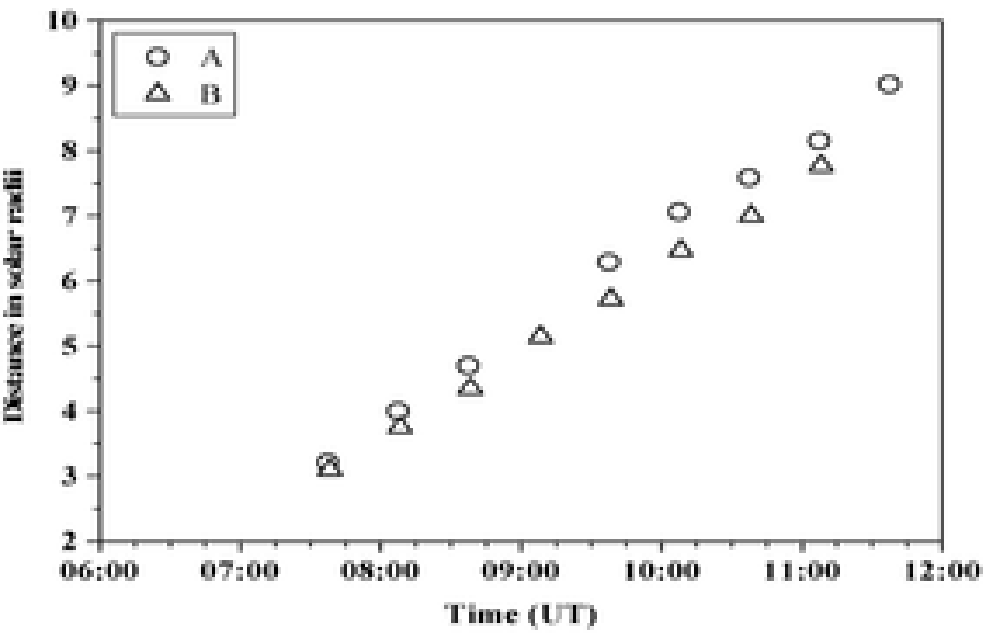}&\includegraphics[width=5.8cm]{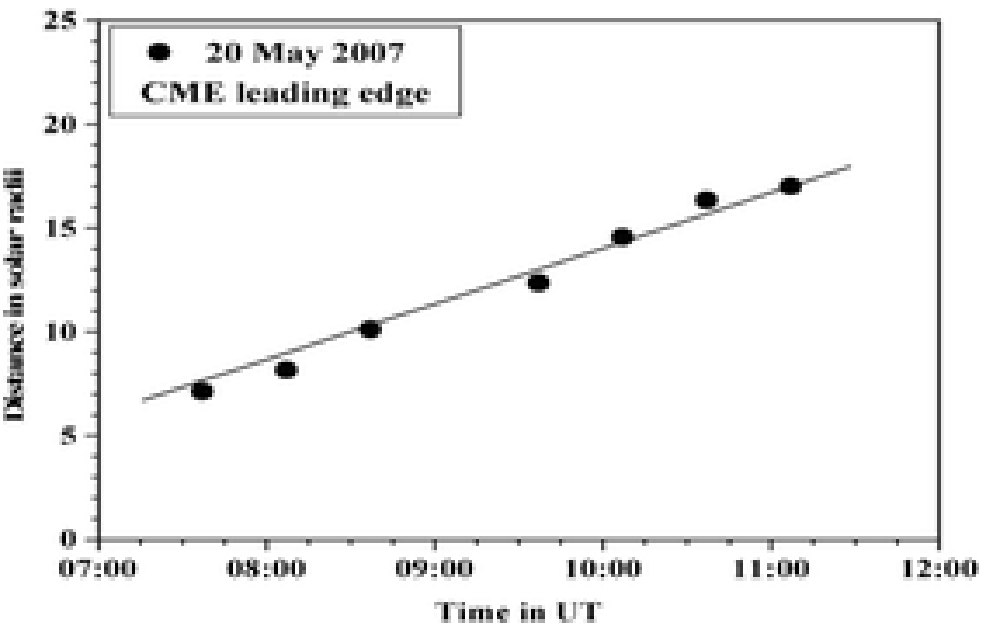}\\
\includegraphics[width=5.8cm]{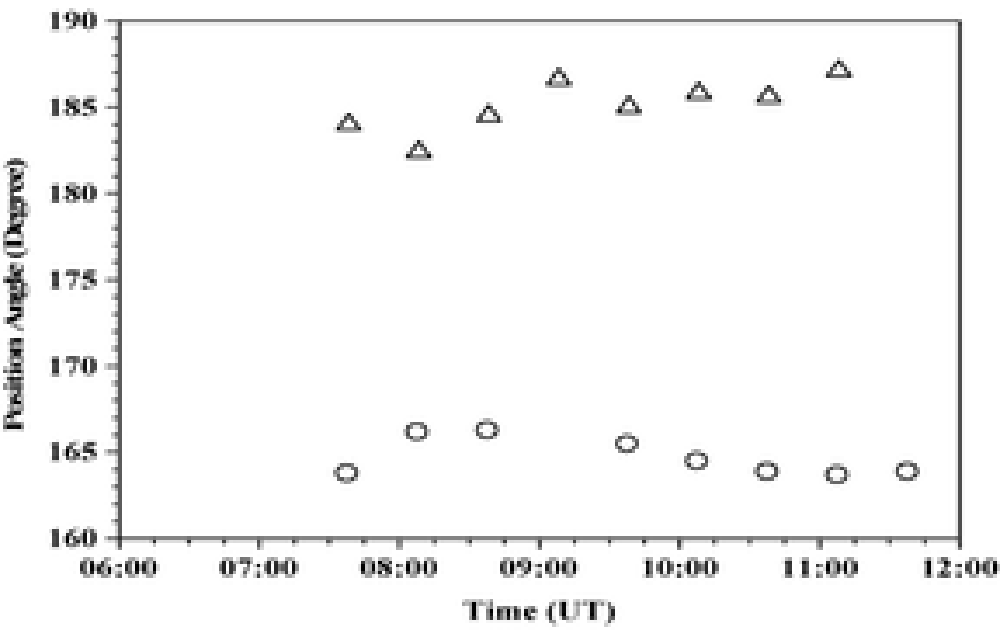} &\includegraphics[width=5.8cm]{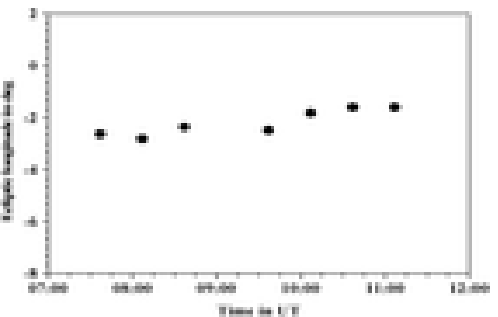}\\
 & \includegraphics[width=5.8cm]{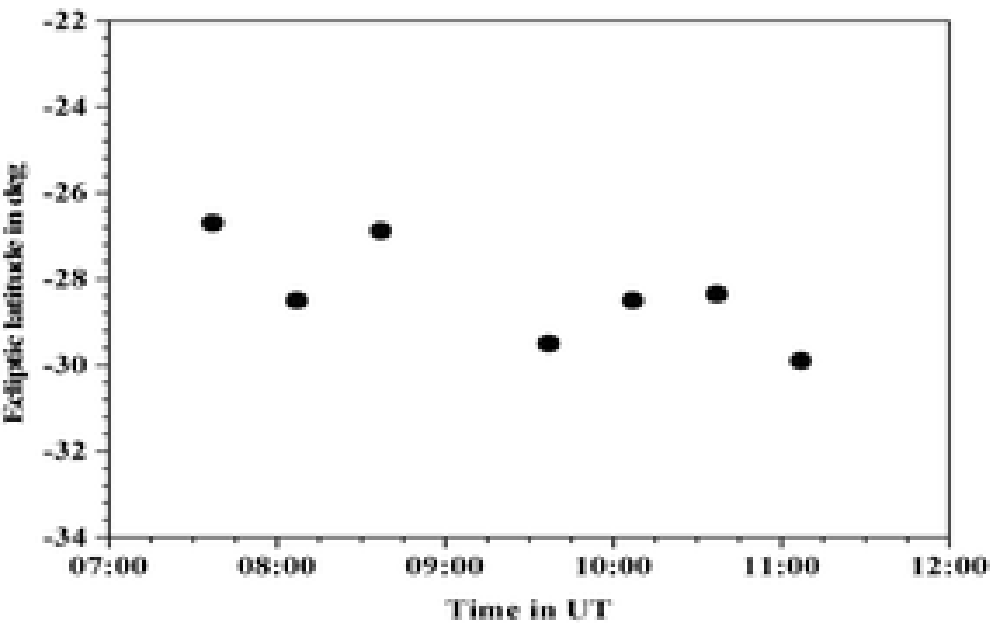}\\
\end{tabular}
\caption[]{\label{n-srivastava-fig:nandita-httime} 
  {\em Left\/}: projected height (top) and position angle
  (bottom) of a selected feature along the leading edge of the 20 May 2007
  CME, as observed in COR2A and B coronagraphs.  {\em Right\/}: 
  true height (top), ecliptic longitude (middle), and ecliptic latitude
  (bottom) for the same feature, obtained using the height-time
  technique for reconstruction.}
\end{figure}



\section{CME observations from SECCHI/STEREO coronagraphs}

The twin STEREO spacecrafts were launched during the solar minimum
period in October 2006, when there was low expectation of the
occurrence of CMEs.  A preliminary study shows that the rate of
CMEs observed with STEREO soon after its launch was 1~CME/day, higher
than the rate of LASCO CMEs recorded during solar minimum. COR1
data show that about 353 CMEs have been recorded until 30 January
2009, as shown at the COR1 website
\url{http://cor1.gsfc.nasa.gov/docs/prelim__events}. The rate decreased
to 0.5~CME/day in 2008, but shows a rising trend since then. Some of
these CMEs could also be tracked in the outer corona with the COR2 and
HI coronagraphs. As the STEREO observations provide simultaneous
images of the corona from two vantage points, \ie\ from ``ahead'' and
``behind'' spacecrafts, they are useful to study the 3-D structure of
CMEs. Prior to the launch of STEREO, different techniques were
employed to derive the 3-D structure of solar features using SoHO data
(\cite{Pizzo04}; \cite{Inhester06}). The CME propagation properties
were also derived by applying a cone model to the LASCO images
(\cite{Zhao02}; \cite{Michalek03}; \cite{Michalek06}). Other
techniques that have been used for 3-D reconstruction are based on
polarization measurements of the white light corona (\cite{Moran04};
\cite{Dere05}). Based on the findings of \citet{Schwenn05} that the
ratio between lateral expansion and radial propagation of CMEs is a
constant, estimations of radial speeds, and hence the arrival time of
CMEs at the Earth, were made.

Recently, with the launch of the twin spacecrafts STEREO A and B, disk
observations of the solar atmosphere in extreme ultraviolet wavelengths
(EUVI) and coronal observations in white light using the SECCHI
coronagraphs from two vantage points simultaneously became
available. This was used to study to study 3-D structure by
reconstruction of solar features such as flare loops and CMEs. Using
stereo pair images, one can also determine the true speeds and
the directions of the leading edge and prominence of a CME. These are
extremely valuable for space weather predictions, as one can not only
estimate the true speeds and propagation direction of a CME in the
corona, but also the exact arrival time at the Earth.  A number
of studies in this direction using STEREO/SECCHI and EUVI images have
been made recently. These studies are mainly based on tie-pointing
reconstruction of STEREO images (\cite{Mierla08}; \cite{Mierla09},
\cite{Srivastava09}). The technique has proven to be extremely
successful when applied to the leading edge of white light CMEs
(\cite{Mierla08}; \cite{Mierla09}) and disk filament and loops
(\cite{Gissot08} and \cite{Aschwanden08}). Essentially, the
tie-pointing technique for reconstructing CMEs is based on epipolar
geometry, wherein the position of the two STEREO spacecrafts A and B
and a point on the solar surface define a plane called the epipolar
plane.  The STEREO mission's plane is a special epipolar plane,
passing through the Sun's centre and the two spacecrafts with its
normal oriented towards the ecliptic North direction. The projections
of all epipolar planes in the spacecraft's images are seen as epipolar
lines in one stereoscopic image that passes through the same epipolar
line in the other stereoscopic image. Tie-pointing involves finding a
one-to one correspondence of a feature in both the stereo images along
equal epipolar lines, calculating the line-of-sight ray that belongs
to the respective images, and eventually constraining the rays to lie
on the same epipolar plane (\cite{Trucco98}).

\begin{table}
\begin{center}
\caption[]{\label{n-srivastava-tbl:reconstruction} Reconstruction of the
leading edge of the 20 May 2007 CME}
 \renewcommand{\arraystretch}{1.2}
\begin{tabular*}{\textwidth}{@{\extracolsep{\fill}}llcccccl}
\hline
 Method& Data & \multicolumn{2}{c}{Projected} & \multicolumn{3}{c}{Reconstruction}& Remarks \\
 & & \multicolumn{2}{c}{$V_{\rm proj}$} &$V_{\rm rec}$ & $\lambda$ & $\theta$ & \\
& & \multicolumn{2}{c}{[km s$^{-1}$]} & [km s$^{-1}$]& [deg] & [deg] &\\
\hline
height-time & COR1 A\&B  & 242 & 253 & 548 & $-2$ & $-27$ & Mierla et al.\ (2008) \\
& COR2 A\&B & 295 &250 & 544 & $-2$ & $-28$ & this paper\\
tie-pointing & COR1 A\&B&    & & 510 & 11 & $-30$ & Srivastava \\
 & COR2 A\&B& & &&& & et al.\ (2009)\\
\hline
\end{tabular*}
\end{center}
\end{table}

A quick method based on tie-pointing is the height-time method for the
3-D reconstruction of CME features (\cite{Mierla08}). This is based on
estimating the projected or ``plane-of-sky'' speeds of selected moving
features of CMEs. \citet{Mierla08} estimated the true heights, speeds
and directions of the leading edges of three CMEs using COR1
coronagraphic images. An example of 3-D reconstruction of one of the
CMEs dated 20 May 2007 studied in their paper is shown in
Fig.~\ref{n-srivastava-fig:mierla-httime}. By applying the
height-time method to COR1 images, \citet{Mierla08} found that the 20
May 2007 CME was located at ecliptic longitude and latitude of around
$-2^{\circ}$ and $-27^{\circ}$ (\ie\ south of the ecliptic),
respectively. The plane-of-sky speeds as measured from the COR1\,A and
COR1\,B coronagraphs were estimated to be approximately 242 and
253~\kms.  The true speed was estimated to be approximately 548~\kms.
In this contribution, we extended the analysis to COR2 images which
cover a field of view from 2--15~R$_{\odot}$. Using the stereo pair
images of the COR2 coronagraphs and applying height-time analysis on
the same feature as in the COR1 images, we obtained the projected
speed as approximately 298 and 250~\kms\ in the A and B images,
respectively (Fig.~\ref{n-srivastava-fig:nandita-httime}). The
reconstructed speed was estimated at 544~\kms. The average ecliptic
longitude and latitude were estimated to be $-2^{\circ}$ and
$-28^{\circ}$ respectively. A comparison of the reconstruction
parameters obtained using different techniques for the leading edge of
the May 20, 2007 CME is given in
Table~\ref{n-srivastava-tbl:reconstruction}. The projected speeds
($V_{\rm proj}$) of the leading edge are given in the third column;
the reconstructed parameters \ie\ the reconstructed speed ($V_{\rm
rec}$), the ecliptic longitude ($\lambda$), and the ecliptic latitude
($\theta$) of the identified feature along the leading edge are given
in the fourth column. The table shows that the results from the
different techniques are in good agreement. This implies that the
leading edge moves at almost constant speed, with only small
deflections in the direction of propagation.

A magnetic cloud was found to be associated with the CME of 20 May
2007 (\cite{Kilpua08}). This cloud arrived at the STEREO A spacecraft at
00:26~UT on May 23. Thus, the actual travel time of the CME to the
Earth was 68 hours. 
In addition, the measured speed of the magnetic cloud at STEREO A was
535~\kms. Assuming that this speed is the average speed at which the
CME travelled towards the Earth, the predicted travel time is about 74
hours, which is equal to the actual arrival time within the
measurement errors. It should be emphasized that use of the projected
speeds would yield large errors in the estimation of the travel
time. Thus, 3-D reconstruction using STEREO pair images has an
important bearing on space weather prediction (\cite{Srivastava09}).

\section{Summary and conclusion}
SECCHI observations have surpassed the capabilities of SoHO
observations. With new views of CMEs obtained by applying
reconstruction techniques to the stereo image pairs, our knowledge of
the 3-D structure and propagation of CMEs will advance considerably.
Also, with reliable estimations of true CME speeds and propagation
directions, the solar community is poised to achieve unprecedented
success in space weather prediction.


\begin{acknowledgement}
  The author acknowledges the help of STEREO/SECCHI consortia for
  providing the data used and presented in this paper. The author also
  thanks M.~Mierla (Royal Observatory of Belgium) and
  B.~Inhester of MPS, Germany, for fruitful discussions on
  CME reconstruction using COR1 and COR2 images.
\end{acknowledgement}

\begin{small}


\begin{thebibliography}{30}
\expandafter\ifx\csname natexlab\endcsname\relax\def\natexlab#1{#1}\fi

 
\bibitem[{{Aschwanden} {et al.} (2008)}]{Aschwanden08}
Aschwanden, M., Wülser, J.-P., Nitta, N. V., Lemen, J. R. 2008, \apj,
679, 827
 
\bibitem[{{Brueckner} {et al.}(1995)}]{Brueckner95}
{Brueckner}, G. E., {Howard}, R. A., {Koomen}, M. J., {Korendyke}, C. M.,
{Michels}, D. J., {Moses}, J. D., et al. 1995, \solphys, 162, 357

\bibitem[{{dal Lago} {et al.}(2004)}] {dal Lago04}
{dal Lago}, A. {Vieira}, L. E. A., {Echer}, E., {Gonzalez} W. D.,
2004, \solphys, 222, 323

\bibitem[{{Delaboudini\'{e}re} {et al.}(1995)}]{Delaboudiniere95}
{Delaboudini\'{e}re}, J.-P., {Artzner}, G. E., {Brunaud}, J., {Gabriel},
A. H., {Hochedez}, J. F., {Millier}, F., {Song}, X. Y., {Au}, B.,
{Dere}, K. P., {Howard}, R. A., and {18 coauthors}, \solphys, 162,
291

\bibitem[{{Dere}, {Wang} \& {Howard} (2005)}]{Dere05}
{Dere}, K. P., {Wang}, X., {Howard}, R. 2005, \apj, 620, L119.

\bibitem[{{Gissot} {et al.}(2008)}]{Gissot08}
{Gissot}, S. F., {Hochedez}, J.-F., {Chainais}, P., {Antoine}, J.-P.,
\solphys, 252,397

\bibitem[{{Gopalswamy} {et al.}(2009)}]{Gopalswamy09}
{Gopalswamy}, N., {Yashiro}, S., {Michalek}, G., {Stenborg}, G.,
{Vourlidas}, A., {Freeland}, S., {Howard}, R. 2009, Earth, Moon
and Planets, doi\,10.1007/s11038-008-9282-7
 
\bibitem[{{Hapgood}(1992)}]{Hapgood92}
{Hapgood}, M. A. 1992, Plan.\ Space Science, 40, 711

\bibitem[{{Howard} {et al.}(2008)}]{Howard08}
{Howard}, R.A., {Moses}, J.D., {Vourlidas}, A., {Newmark}, J.S.,
{Socker}, D.G., {Plunkett}, S.P., et al.\ 2008, Space Sci. Rev., 136,
67

\bibitem[{{Hudson} {et al.} (2007)}]{2007ISSI}
{Hudson} et al. in Coronal Mass Ejections eds. H. Kunow, N.U. Crooker,
J.A. Linker, R. Schwenn, \& R. Von Steiger, Space Science Series of
ISSI 2007, 13

\bibitem[{{Inhester}(2006)}]{Inhester06}
{Inhester}, B. 2006, Publ.\ Int.\ Space Sci.\ Inst., astro-ph/0612649,
in press

\bibitem[{{Kaiser} {et al.}(2008)}]{Kaiser08}
{Kaiser}, M. L., {Kucera}, T. A., {Davila}, J. M., {St. Cyr}, O. C.,
{Guhathakurta}, M., {Christian}, E. 2008, Space Sci.\ Rev., 136, 5.

\bibitem[{{Kilpua} {et al.}(2008)}]{Kilpua08}
{Kilpua}, E. K. J., {Liewer}, P. C., {Farrugia}, C., {Luhmann}, J. G.,
{Moest}, C., {Li}, Y., {Liu}, Y., {Lynch}, B. J., {Russell}, C. T.,
{Vourlidas}, A., {Acuna}, M. H., {Galvin}, A. B., {Larosn}, D.,
{Sayvaud}, J. A. 2008, \solphys, in press

\bibitem[{{Michalek} {et al.}(2003)}]{Michalek03}
{Michalek}, G., {Gopalswamy}, N., {Yashiro}, S. 2003, \apj, 584, 472

\bibitem[{{Michalek}(2006)}]{Michalek06}
{Michalek}, G. 2006, \solphys, 237, 101

\bibitem[{{Mierla} {et al.}(2008)}]{Mierla08}
{Mierla}, M. {Davila}, J., {Thompson}, W., {Inhester}, B.,
{Srivastava}, N., {Kramar}, M., {St. Cyr}. O. C., {Stenborg}, G,
{Howard}, R. A. 2008,\solphys, 252, 385

\bibitem[{{Mierla} {et al.}(2009)}]{Mierla09}
{Mierla}, M., {Inhester}, B., {Marque}, C., {Rodriguez}, L., {Gissot},
S., {Zhukov}, A., {Berghmans}, D., {Davila}, J. 2009, \solphys,
submitted

\bibitem[{{Moran} \& {Davila}(2004)}]{Moran04}
{Moran}, T.G., {Davila}, J. 2004, Science, 305, 66

\bibitem[{{Pizzo} \& {Biesecker}(2004)}]{Pizzo04}
{Pizzo}, V.J., {Biesecker}, D.A. 2004, Geophys.\ Res.\ Lett., 31, L21802

\bibitem[{{Schwenn} {et al.}(2005)}]{Schwenn05}
{Schwenn}, R, {Dal Lago}, A., {Huttunen}, E., {Gonzalez}, W.D.,
2005, Ann.\ Geophys., 23, 1033

\bibitem[{{Schwenn} {et al.}(2007)}]{Schwenn07}
{Schwenn} et al. in Coronal Mass Ejections eds. H. Kunow,
N.U. Crooker, J.A. Linker, R. Schwenn, \& R. Von Steiger, 2007,
Springer, Berlin, 137

\bibitem[{{Srivastava} {et al.}(1999)}]{Srivastava99}
Srivastava, N., Schwenn, R., Inhester, B., Stenborg, G., \& Podlipnik,
B. 1999, in Solar Wind 9, AIP Conf.\ Proc.\ 471, 115

\bibitem[{{Srivastava} {et al.}(2000)}]{Srivastava00}
{Srivastava}, N., {Schwenn}, R., {Inhester}, B., {Martin}, S. F.,
{Hanaoka}, Y. 2000, \apj, 534, 468

\bibitem[{{Srivastava} {et al.}(2009)}]{Srivastava09}
Srivastava, N., Inhester, B., Mierla, M., Podlipnik, B. 2009,
\solphys, submitted

\bibitem[{{St. Cyr} {et al.}(2000)}]{StCyr00}
{St. Cyr}, O. C., {Howard}, R. A., {Sheeley}, N. R., {Plunkett},
S. P., {Michels}, D. J., {Paswaters}, S. E., {Koomen}, M. J.,
{Simnett}, G. M., {Thompson}, B. J., {Gurman}, J. B., {Schwenn}, R.,
{Webb}, D. F., {Hildner}, E., {Lamy}, P. L. 2000, J.~Geophys.~Res.,
105, 18169

\bibitem[{{Thompson} {et al.}(2003)}]{Thompson03}
{Thompson}, W. T., {Davila}, J. M., {Fisher}, R. R., {Orwig}, L. E.,
{Mentzell}, J. E., {Hetherington}, S.E. 2003, in {Keil}, S. L.,
{Avakyan}, S. V. (eds.), Innovative Telescopes and Instrumentation for
Solar Astrophysics, SPIE, 4853, 1

\bibitem[{{Thompson}(2006)}]{Thompson06}
{Thompson}, W.T. 2006, \aap, 449, 791

\bibitem[{{Tousey}(1973)}]{Tousey73} 
{Tousey} R., The Solar Corona, in Space Research XIII 1973,
eds. M. J. Rycroft, S. K. Runcorn, 713, Akademie-Verlag, Berlin

\bibitem[{{Trucco} \& {Verri}(1998)}]{Trucco98}
{Trucco}, E., {Verri}, A. 1998, Introductory Techniques for 3-D
Computer Vision, Prentice Hall

\bibitem[{{Vourlidas} {et al.}(2002)}]{Vourlidas02}
{Vourlidas}, A., {Buzasi}, D., {Howard}, R. A., {Esfandiari}, E. in:
Solar variability: from core to outer frontiers. The 10th European
Solar Physics Meeting, ed.\ A. Wilson, ESA SP-506, 91

\bibitem[{{Wang} {et al.}(2002)}]{Wang02}
{Wang}, Y. M., {Ye}, P. Z., {Wang}, S., {Zhou}, G. P., {Wang},
J. X. 2002, J.~Geophys.~Res., 107, 2

\bibitem[{{Xie}, {Ofman} \& {Lawrence}(2004)}]{Xie04}
{Xie}, H., {Ofman}, L., {Lawrence}, G. 2004, J.~Geophys.~Res.,
109, A03109.

\bibitem[{{Yashiro} {et al.} (2004)}]{Yashiro04}
{Yashiro}, S., {Gopalswamy}, N., {Michalek}, G., {St. Cyr}, O. C.,
{Plunkett}, S. P., {Rich}, N. B., {Howard}, R. A. 2004,
J.~Geophys.~Res., 109, A07105.

\bibitem[{{Yurchshyn} {Wang} \& {Abramenko}(2003)}]{Yurchshyn03}
Yurchyshyn, V., Wang, H., Abramenko, V. 2003, Adv.\ Sp.\ Res., 32,
1965

\bibitem[{{Zhang} {et al.}(2003)}]{Zhang03}
{Zhang}, J., {Dere}, K. P., {Howard}, R. A., {Bothmer}, V. 2003,
\apj, 582, 520

\bibitem[{{Zhang} {et al.}(2004)}]{Zhang04}
{Zhang}, J., {Dere}, K. P., {Howard}, R. A., {Vourlidas}, A.,
2004, \apj, 604, 420

\bibitem[{{Zhao}, {Plunkett} \& {Liu}(2002)}]{Zhao02}
{Zhao}, X. P., {Plunkett}, S. P., {Liu}, W. 2002, J.~Geophys.~Res.,
107, 1223

\bibitem[{{Zhao} \& {Webb} (2003)}]{Zhao03}
{Zhao}, X. P., {Webb}, D. F. 2003, J.~Geophys.~Res., 108, 4

\end{thebibliography}


\end{small}

\end{document}